\documentclass[10pt, a4paper, twocolumn]{article}

\usepackage[english]{babel}
\usepackage[numbers]{natbib}
\usepackage{bigfoot}
\usepackage{algorithmic}
\usepackage{array}
\usepackage{mdwmath}
\usepackage{mdwtab}
\usepackage{eqparbox}
\usepackage{stfloats}
\usepackage{url}
\usepackage{textcomp}
\usepackage{graphicx}
\usepackage{subcaption}
\usepackage{booktabs}

\usepackage[usenames, dvipsnames]{color}
\usepackage{comment}

\usepackage[noblocks,affil-it]{authblk}
\usepackage[hmargin=0.5in,vmargin=0.9in]{geometry}

\begin{document}

\title{Cobalt: \\ A GPU-based correlator and beamformer for LOFAR}



\author[1]{P.\,Chris Broekema\thanks{Corresponding author: \texttt{broekema@astron.nl}}}
\author[1]{J.\,Jan David Mol}
\author[1]{Ronald Nijboer}
\author[1]{Alexander\,S. van Amesfoort}
\author[1]{Michiel\,A. Brentjens}
\author[1]{G.\,Marcel Loose}
\author[1]{Wouter\,F.\,A. Klijn}
\author[1]{John\,W. Romein}
\affil[1]{\small{ASTRON, the Netherlands Institute for Radio Astronomy, Postbus 2, 7990 AA, Dwingeloo, the Netherlands}}

\date{}
\maketitle

\begin{abstract}
For low-frequency radio astronomy, software correlation and beamforming on general purpose hardware is a viable alternative to custom designed hardware.
LOFAR, a new-generation radio telescope centered in the Netherlands with international stations in Germany, France, Ireland, Poland, Sweden and the UK, has successfully used software real-time processors based on IBM Blue Gene technology since 2004.
Since then, developments in technology have allowed us to build a system based on commercial off-the-shelf components that combines the same capabilities with lower operational cost.
In this paper we describe the design and implementation of a GPU-based correlator and beamformer with the same capabilities as the Blue Gene based systems.
We focus on the design approach taken, and show the challenges faced in selecting an appropriate system.
The design, implementation and verification of the software system shows the value of a modern test-driven development approach.
Operational experience, based on three years of operations, demonstrates that a general purpose system is a good alternative to the previous supercomputer-based system or custom-designed hardware.
\end{abstract}

\section{Introduction}
\label{sec:introduction}

The LOw Frequency ARray (LOFAR)~\citep{vanHaarlem:2013dsa} radio telescope is often described as one of the first of a new generation of software telescopes.
LOFAR has pioneered the use of a combined software correlator and beamformer in an operational radio telescope since 2004~\citep{Romein:06,Romein:10,Mol:11}.
One key characteristic of a software telescope is the ability to ride the technology wave to increase functionality and/or reduce operational cost by leveraging new developments.
In this paper we discuss the hardware design of the third generation Graphics Processing Unit (GPU) based LOFAR software correlator and beamformer: Cobalt (\textbf{CO}rrelator and \textbf{B}eamformer \textbf{A}pplication for the \textbf{L}OFAR \textbf{T}elescope), as well as the design and development of the associated software.

Since the tasks of this real-time central processor are well known and clearly defined, this application is an excellent candidate for a focused hardware/software co-design approach.

In this paper we describe the following concepts that in combination led to the success of Cobalt:
\begin{itemize}
\item A data flow-driven design philosophy;
\item Hardware/software co-design; 
\item Data flow analysis and task mapping to identify potential weaknesses in available HPC solutions for our streaming application;
\item Close public-private collaboration in the hardware design, which showed the clear advantages of such a partnership in this kind of project;
\item A simplified system engineering approach in the design and implementation phases of the project;
\item An agile software engineering methodology to ensure timely delivery within budget;
\item A Test-driven software development process to improve the robustness of our system.
\end{itemize}

\section{Related work}
\label{sec:related}
The Cobalt project built on previous experience with combined software correlator and beamformer systems in the LOFAR telescope~\citep{Romein:06,Romein:10,Mol:11}.
We discuss some aspects of these in more detail in Section \ref{sec:rtcp}.
Cobalt shared common ancestry with the AARTFAAC correlator~\citep{prasad:16}, although the radically different I/O ratio led to different design decisions.

There are several other software correlators in use in radio astronomy.
Here we briefly discuss some of these in relation to the Cobalt system.
We limit ourselves to FX-correlators, that combine a filter- and Fourier transform (F) stage with a cross-correlation (X) stage

The correlators used by the Murchison Widefield Array (MWA)~\citep{oord:2015}, the Large Aperture Experiment to Detect the Dark Ages (LEDA)~\citep{Kocz:15}, and PAPER~\citep{Parsons:10} all share the same general architecture.
Whereas Cobalt implements both the filter (F-stage) and the correlator (X-stage) in GPUs, the above mentioned instruments employ a hybrid FPGA-GPU approach.
The F-stage is implemented in FPGA, the X-stage is implemented in GPUs using the xGPU library~\citep{Clark:11}.
A high bandwidth switch connects the F- and X-stages.


The Giant Metrewave Radio Telescope (GMRT) real-time software backend~\citep{gupta:2010} uses a structure similar to the MWA correlator, with nodes dedicated to three specific tasks.
In this case the software backend relies on conventional CPUs only, with heavy use of off-the-shelf performance optimized libraries.

For Very Long Baseline Interferometry (VLBI) a number of software correlators have been developed.
Examples of these are SFXC, developed by JIVE~\citep{Keimpema:15}, and DiFX~\citep{Deller:07,Deller:11}.
These perform tasks similar to Cobalt, although DiFX does not include a beamformer.
However, data rates are usually modest compared to those generated by the LOFAR stations.

A real-time software correlator has been developed and deployed for the Canadian Hydrogen Intensity Mapping Experiment (CHIME) pathfinder~\citep{Denman:15}.
The correlator stage of this system is  very similar in concept and size to Cobalt, but implements the F-stage in FPGAs, requiring additional communication from the F to the X stage.
In Cobalt the F and X stage use the same hardware.

\section{LOFAR: the Low Frequency Array}
\label{sec:lofar}
LOFAR, the LOw Frequency ARay, is a new-generation radio telescope, built in the northern part of the Netherlands, with international stations distributed across Europe.
As the name suggests, it is designed to observe in the relatively low and unexplored frequency range of 10 to 250\,MHz.
The array consists of 40 stations: 24 core and 16 remote, in the Netherlands, with an additional 13 international stations, 6 in Germany, 3 in Poland and one each in France, Ireland, Sweden, and the UK.

Each station consists of two receiver types, low band dipole antennas and high band antenna tiles, covering either side of the commercial FM band.
A LOFAR station consists of 96 Low Band Antennas (LBAs), operating from 10 to 90\,MHz.
In addition, Dutch core stations have 48 High Band Antenna (HBA) tiles in two clusters that cover the frequency range from 110 to 250\,MHz.
Remote stations in the Netherlands have the same number of HBA tiles, in a single cluster.
International stations provide a single cluster of 96 HBA tiles.

At each LOFAR station dedicated processing equipment samples, digitises and digitally filters data using a polyphase filter bank.
This filterbank produces 512 frequency bands with a spectral bandwidth of 195\,kHz.
By coherently adding the same frequency bands of  individual antennas or tiles, station beams are created.
Such a beamformed frequency block is referred to as a \textit{subband}, 488 of which may be selected per observation, giving a total spectral bandwidth of 95\,MHz (in the most common 8-bit mode).
Spectral bandwidth may be exchanged for beams, essentially allowing up to 488 independent (narrow-band) pointings to be made.
Core stations may be split, allowing the two HBA clusters to be treated as smaller, but fully independent, stations.
To distinguish them from  physical stations, these are called \textit{antenna fields}, 77 of which currently make up the LOFAR array.
Subbands produced by these antenna fields are transported to the central processing facility, hosted by the University of Groningen, about 50 kilometers from the LOFAR core area, using UDP/IP over many 10\,Gigabit Ethernet links.

The central processor can be divided into three distinct components: the real-time processor, the post-processing cluster, and the archive.
The real-time processor (Cobalt), which implements a correlator and beamformer, is a soft real-time system that collects data from the antenna fields, conditions this data, applies a second polyphase filter, and subsequently combines all antenna fields (beamformer mode) or all antenna field pairs (correlator mode) to produce intermediate results.
Although there is no hard deadline in the sub-second range as in a classic real-time system, it is required that the central processor keeps up with the antenna field data streams, otherwise data is irretrievably lost.
In Section \ref{sec:processing-steps} we discuss the processing steps that make up the real-time system.

Output from the real-time processor is stored on the post-processing cluster.
This is a conventional Linux cluster, with significant disk capacity to store intermediate products and facilitate further processing.
Here, instrument calibration is performed, possible interference is identified and removed, and final products (images, pulse profiles, source lists) are created.

\begin{figure*}[htb]
\centering \includegraphics[width=0.8\textwidth]{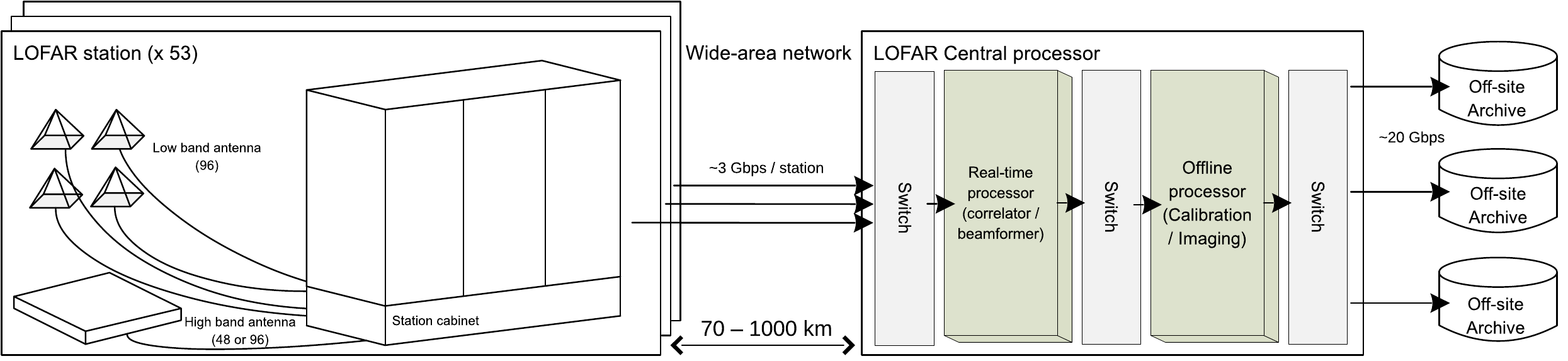}
\caption{Top-level overview of the LOFAR system.}
\label{fig:lofar-top-level}
\end{figure*}

Final products are exported to the LOFAR long-term archive, which is currently distributed over three sites: Amsterdam hosted by SURFsara in the Netherlands, J\"ulich hosted by Forschungszentrum J\"ulich in Germany, and Poznan hosted by PSNC in Poland.
Astronomers retrieve their data from one of these archive sites, no end-user interaction with the LOFAR system is required.
Figure \ref{fig:lofar-top-level} shows a top-level overview of the LOFAR system.

The remainder of this paper will focus on the LOFAR real-time processor.

\subsection{The LOFAR real-time processor}
\label{sec:rtcp}
The LOFAR project decided early on to employ general purpose computing for the real-time processor, both to exploit the fact continued developments in general purpose processor technology had made this feasible, and to save precious FPGA development resources for the station processing boards.
Initially, the requirements for the LOFAR real-time processor were quite challenging for a general purpose compute system. 
The only feasible option available was to use a supercomputer.
In 2004, an IBM Blue Gene/L was installed at the LOFAR central processor.
At the time, the LOFAR real-time processor was the fastest supercomputer in the Netherlands, and the second fastest in Europe\footnote{\url{https://www.top500.org/lists/2005/06/}}.
Although compute performance of the Blue Gene/L was sufficient, significant research and development was required to achieve the required I/O performance~\citep{Romein:06, Iskra:08}.

In 2008 the six rack Blue Gene/L system was upgraded to a slightly more powerful, but much smaller and more energy-efficient three rack Blue Gene/P system.
While a significant improvement over its predecessor in terms of programming environment and general hardware features, considerable research was again required to achieve the I/O performance required~\citep{Romein:09a, Kazutomo:09, Kazutomo:10}.

While the Blue Gene real-time processors were operational, research continued into various other software alternatives~\citep{Romein:10, Nieuwpoort:11}.
The advent of many-core architectures for high performance computing, in particular Graphics Processing Units (GPUs), allowed us to move away from supercomputers, and instead build the third-generation correlator and beamformer based on general-purpose server hardware and accelerators.
In this paper we discuss the design approach taken, we show some of the problems encountered and how these were tackled, and we conclude with the successful commissioning into operational service of a new, GPU-based, LOFAR correlator and beamformer.
Several years of operational statistics are presented in Section \ref{sec:operational}.

\subsection{Processing steps}
\label{sec:processing-steps}
The LOFAR real-time processor receives data from LOFAR antenna fields as continuous UDP/IP data streams. 
Missing and out-of-order packets are identified and, where possible, corrected.
Data that has not arrived after a short deadline is considered lost.
Each incoming data stream contains all frequency data from a single antenna field, while each processing node for a given frequency range requires data from all antenna fields.
Therefore, a transpose is required on the incoming data, before further processing.

\begin{figure*}[hbt]
\centering \includegraphics[width=0.8\textwidth]{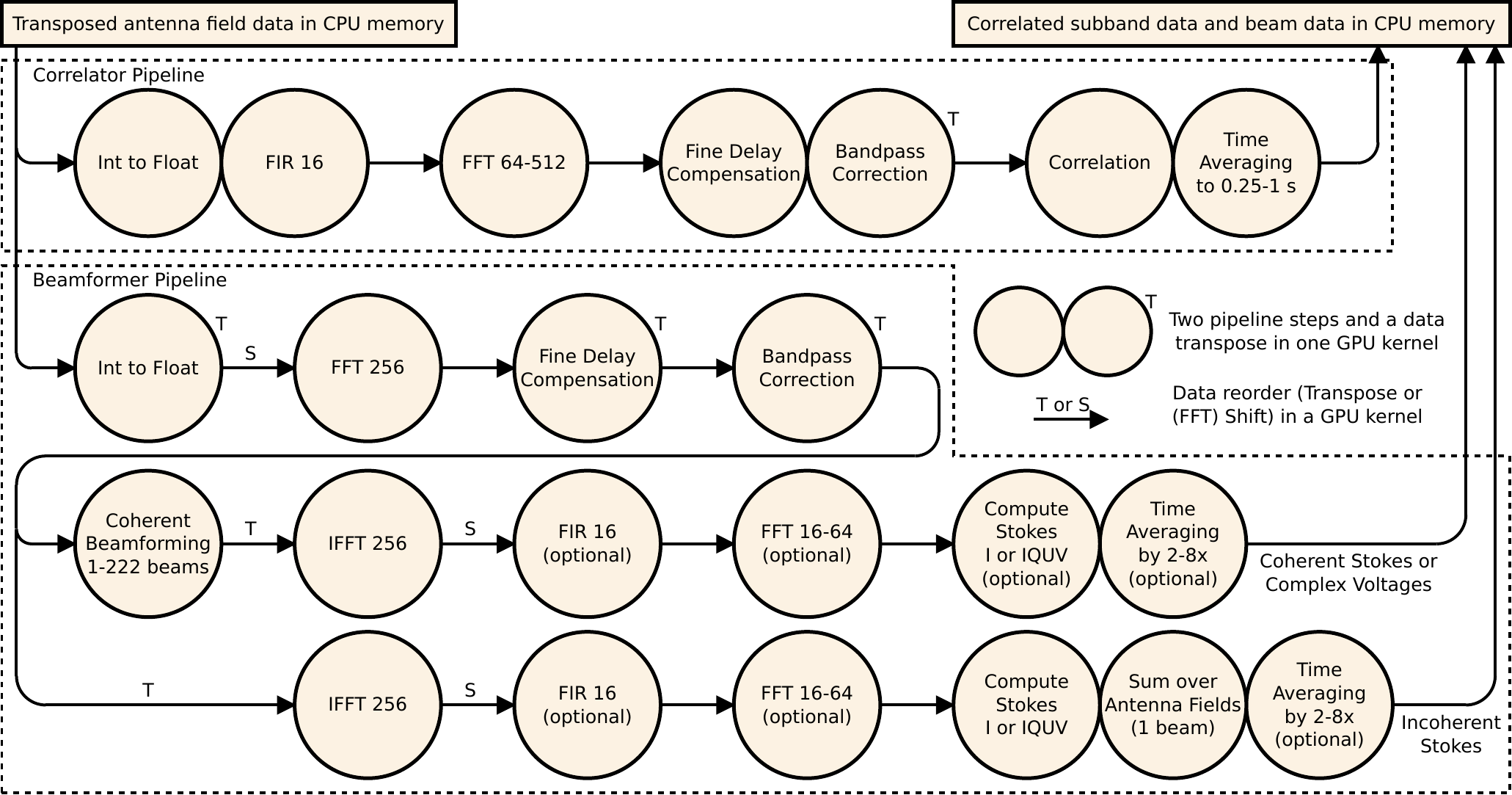}
\caption{Signal processing steps in the correlator and beamformer pipelines.}
\label{fig:signal-processing}
\end{figure*}

The processing component is complex and involves a number of optional sub-components, as shown in Figure \ref{fig:signal-processing}.
Data is converted from fixed to floating point to better match the hardware available in a general purpose computer.
The current LOFAR real-time processor uses single precision complex floating point throughout, with one exception: calculating delays for delay compensation.
Two main pipelines are implemented that can run in parallel.
The correlator pipeline implements an FX-style correlator, the components of which were described in more detail in earlier work~\citep{Romein:10}. 
The beamformer pipeline consist of coherent and incoherent components, as well as a complex voltage pipeline, details of which were previously published as well~\citep{Mol:11}.
In Section \ref{sec:gpu_kernels} we describe how these processing steps are implemented in Cobalt.

\section{Development process}
The relatively modest scale of the project allowed us to use a slightly simplified system engineering approach in the design of this system. 
First, the system requirements, both functional and non-functional, were identified (see Section \ref{sec:system-requirements}).
From these, a high-level architecture was derived (Section \ref{sec:hardware-design}).
This, combined with a detailed analysis of the various aspects of the application performance profile, such as network data flow (Section \ref{sec:dataflow}), memory footprint (Section \ref{sec:memory-bandwidth}) and computational load (Section \ref{sec:requirements} and Section \ref{sec:accelerator}) led to a detailed design of the system hardware (Section \ref{sec:cobalt-hw}).
During the hardware implementation phase, a single prototype node was used to verify that the selected hardware implementation met the performance requirements (Section \ref{sec:prototyping}).
Finally, the fully deployed system was verified against the system requirements (Section \ref{sec:verification}).
This process closely mirrors a traditional systems engineering approach, but the relatively small project size meant we could simplify the process by eliminating most of the formal documentation.

Before the start of the Cobalt project the feasibility of a GPU-based solution was researched and optimized GPU kernels had been developed.
Moreover, a highly optimized and proven Blue Gene implementation of all required functionality was available.
However, a different hardware architecture and steep performance, maintenance and reliability requirements, necessitated redesign of our on-line processing software, except for wrapper and support libraries.
The beamformer pipeline was redesigned, so several GPU kernels had to be adapted or rewritten.

The development process was paramount to obtain correct output and adequate performance within a limited time frame.
We used the Agile/Scrum development process~\citep{schwaber:97, Schwaber:2001} to focus a small software team on a common goal, divide and plan remaining work, and periodically tried to improve our practices.

\section{System requirements}
\label{sec:system-requirements}
The following hard and soft requirements were put on the Cobalt system.
In terms of functional requirements, the Cobalt system \emph{must}:
\begin{itemize}
\item be able to correlate 64 antenna fields in 16 bit mode and in 8 bit mode at full bandwidth (for a single beam), i.e.\ with 244 resp.\ 488 subbands, down to 1 sec time resolution and at maximum 256 channels per subband frequency resolution. In this mode up to 8 independent beams can be made, in which case the number of beams, the total bandwidth and the number of bits have to be traded against each other.
\item be able to create 127 time domain data streams using all 48 core antenna fields in 16-bit mode at full bandwidth. These can be recorded in one of three modes: 1) coherent addition (Stokes I only or Stokes IQUV, referred to as a coherent tied-array beam), 2) incoherent addition (Stokes I only or Stokes IQUV, referred to as an incoherent tied-array beam), or 3) coherent complex voltage (XX, XY, YX, YY). Time resolution can be traded against frequency resolution, within the resolution of a subband.
\end{itemize}
In addition, there were the following non-functional requirements.
The system must be delivered in time and within budget.
It must have hardware, software, and data input/output connections installed, tested, and debugged in a staged approach.
The system must have a design that allows to scale up and be prepared for future planned modes and functionality.
Furthermore, the system must have an operational availability greater than 95\% (excluding planned service), while having a system maintenance staff effort of less than 0.25 FTE, delivered during business hours only.
The total operating costs per year must be lower than 50\% of the (one-time) capital investment costs.

The non-functional requirements on Scalability, Operational Availability (i.e.\ robustness) and Maintenance Effort (i.e.\ maintainability) translated into software quality, programming environment, software support, test environment, non-monolithic design, etc.

In addition to the hard requirements there were the following soft requirements, i.e.\ nice-to-haves.
The Cobalt system \emph{should} be able to handle more LOFAR data, such as e.g.\ doing parallel LBA and HBA observing (doubling the number of available subbands, and doubles the required Cobalt capacity for a given observation), correlating more antenna fields (up to 80), correlating longer baselines (up to 3500 km), creating more beams (up to 200), or operating in 4-bit mode (which would double the number of available subbands, at the cost of some dynamic range, again doubling required Cobalt capacity for a given observation).

It should have the capacity to handle additional online tasks, e.g.\ Fly's-eye mode in which we store antenna field data without central beamforming, handling of more than 8 independent beams in correlator mode, online flagging, and beamforming the six central stations (''superterp'') before correlation. 
Cobalt should also have the capacity to handle additional offline processing tasks including automatic flagging, (self-) calibration and averaging, coherent de-dispersion of pulsar data and production of dynamic spectra and additional parts of the pulsar pipeline: online folding and online searching.

Finally, Cobalt should prepare for future extensions, such as commensal observing, parallel observing with sub-arrays, responsiveness to triggers and interrupts, and additional observing modes.

\section{Hardware design and implementation}
\label{sec:hardware-design}

In the design process we focused on data flow rather than compute requirements.
One of the key characteristics of the LOFAR real-time processor is a relatively high data rate.
While making sufficient compute capacity available was a key requirement, efficient use of this capacity critically depends on efficient data flow through the system.
Furthermore, previous many-core correlator research meant that the computational requirements and challenges were relatively well understood.
We therefore made the conscious, if somewhat counter-intuitive, decision to focus our hardware design for the Cobalt system on data flow, with computational capacity a crucial but secondary design goal.

\subsection{Hardware requirements and design priorities}
\label{sec:requirements}
Cobalt was intended to be a drop-in replacement for the existing Blue Gene based correlator and beamformer for LOFAR.
As such, the primary requirement for Cobalt was to provide performance equal to the previous system.
In addition, the desire to increase the number of antenna fields that can be correlated from 64 to 80 was expressed.
Table \ref{tab:requirements} summarizes the top-level hardware requirements for Cobalt.

\begin{table*}[hbt]
  \centering
  \resizebox{\textwidth}{!} {
    \begin{tabular}{lcc}
      \toprule
      Component        & Requirement                                  & Design target \\
      \cmidrule(r){2-3}
      Input  bandwidth & $\sim$192 Gbps (64 antenna fields)           & $\sim$240 Gbps (80 antenna fields)  \\
      Output bandwidth & 80 Gbps                                      & $\geq$80 Gbps                          \\
      Interconnect     & input + output bandwidth                     & $>$2 * (input + output bandwidth)   \\
      Correlator       & 64 antenna fields, 244~(16-bit) - 488~(8-bit) subbands          & 80 antenna fields, 244~(16-bit) - 488~(8-bit) subbands \\ 
      Beamformer      & 127 beams, 48 antenna fields, 244~(16-bit) subbands & 200 beams, 48 antenna fields, 244~(16-bit) subbands \\
      \bottomrule
    \end{tabular}
  }
  \caption{Top-level hardware requirements for Cobalt}
  \label{tab:requirements}
\end{table*}

In order to translate these requirements into a system design, we estimated the required compute capacity.
The main contributors were expected to be:
\begin{itemize}
\item correlator
\item polyphase filter bank (essentially many Finite Impulse Response (FIR) filters, feeding into a Fast Fourier Transform (FFT))
\item bandpass and clock corrections
\end{itemize}

\begin{figure*}[hbt!]
  \centering
  \begin{subfigure}[t]{0.47\textwidth}
    \centering
    \includegraphics[width=\textwidth]{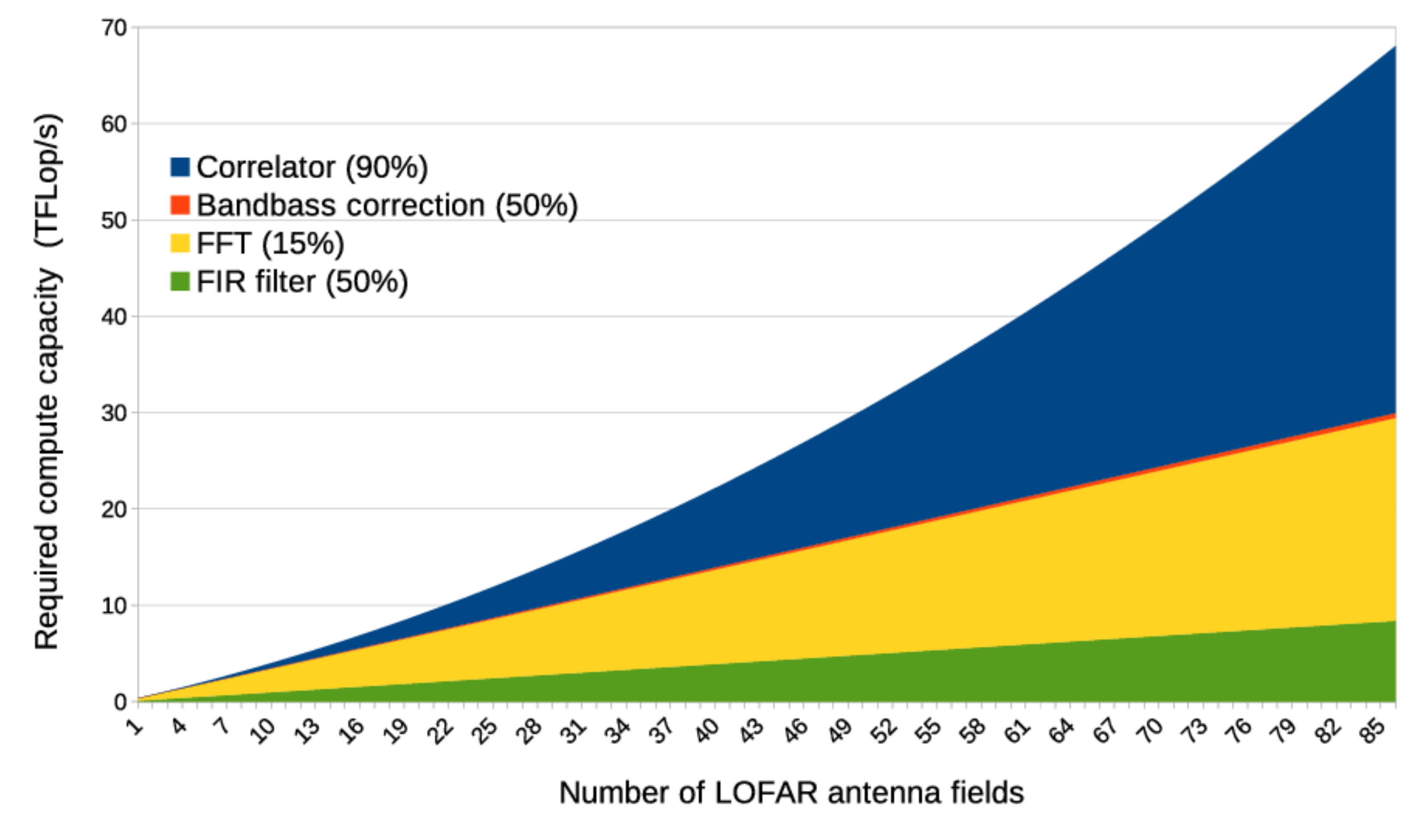}
    \caption{Predicted compute scaling (assumed computational efficiency in parentheses).}
    \label{fig:predicted-performance}
  \end{subfigure}
  \begin{subfigure}[t]{0.47\textwidth}
    \centering
    \includegraphics[width=\textwidth]{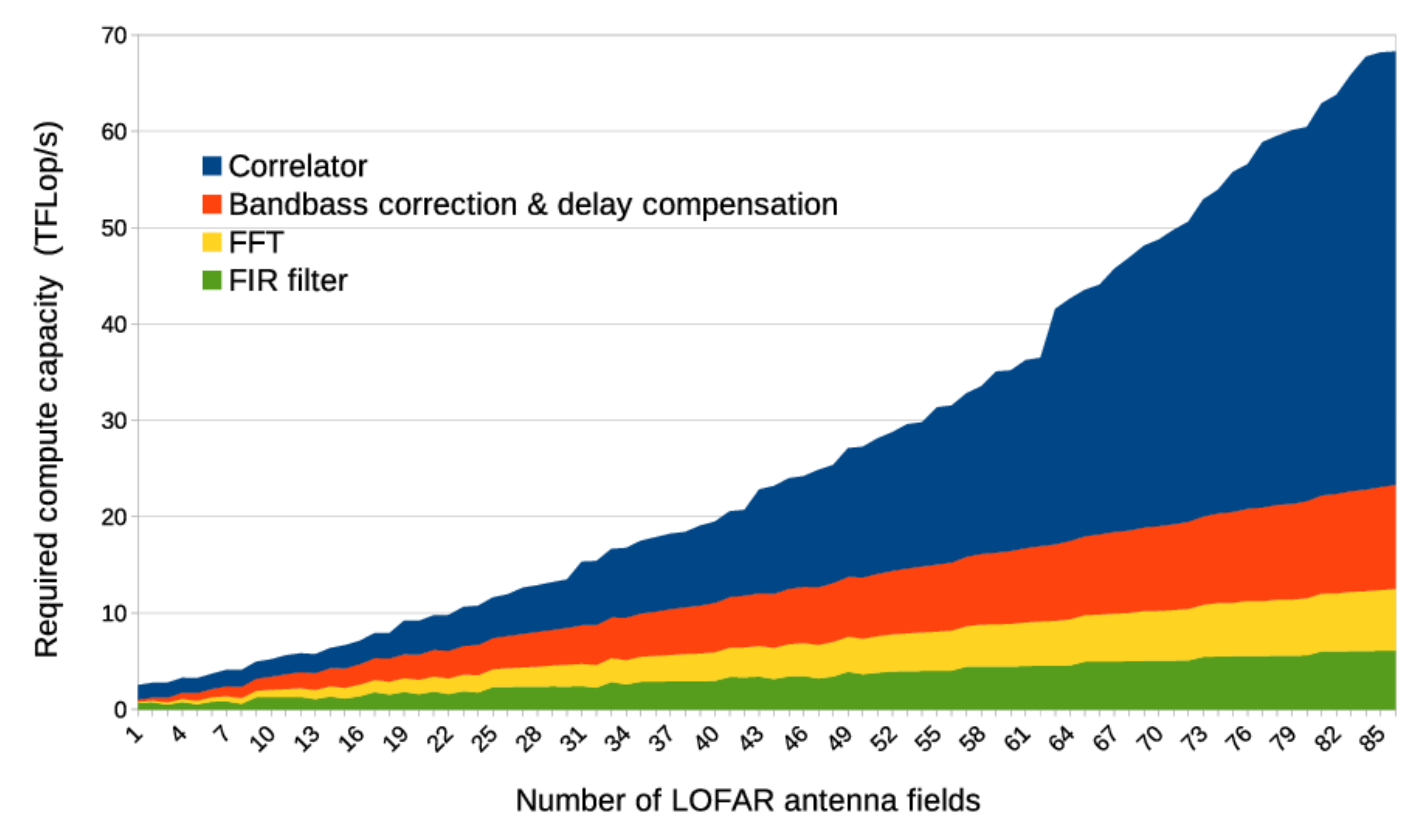}
    \caption{Measured compute performance (November 2017).}
    \label{fig:measured-performance}
  \end{subfigure}
  \caption{Predicted and measured scaling of required compute capacity against number of processed LOFAR antenna fields.}
  \label{fig:predicted-compute-requirements}
\end{figure*}

Figure~\ref{fig:predicted-performance} shows how we expected the compute load to scale with the number of LOFAR antenna fields.
This figure takes the theoretical compute load of each of the contributors, and takes into account a rough estimate of the achievable computational efficiency\footnote{Computational efficiency is defined as the percentage of the theoretically peak performance that is obtained in practice.}.
Extensive prototyping, as well as experience with previous software correlators, showed that the correlator itself can be highly optimized~\citep{Romein:06,Romein:10,Nieuwpoort:11}.
Computational efficiencies well above 90\% of theoretical peak performance have been observed.
The FFT on the other hand is notoriously inefficient.
So while the computational complexity of the correlator is $\mathcal{O}(N^2)$, compared to $\mathcal{O}(N \log N)$ for the FFT\footnote{To complicate matters further, note that in these complexity measures $N$ may not necessarily refer to the same parameter.}, the effective contribution to the computational load of both is much closer than these theoretical computational complexities suggest.
For the purposes of this estimate, we assumed a computational efficiency of 90\% for the correlator and 15\% for the FFT, based on the published performance of the CuFFT library.
The other components were not expected to contribute much to the required compute resources, therefore a computational efficiency of 50\% for all other contributors was used.

In Figure~\ref{fig:measured-performance} we show measured performance scaling of the operational Cobalt system.
While the total consumed resources are very close to the estimate in Figure~\ref{fig:predicted-performance}, there are some marked differences.
The cost of the FFT was significantly overestimated, due a the conservatively chosen complexity.
Both correlation and FIR filters are close to the estimate, but bandpass correction consumes much more compute resources than estimated.
This is due to the addition of delay compensation, compensating earth rotation, but more importantly due to a performance regression discussed in more detail in Section~\ref{sec:verification}.
However, Figure~\ref{fig:measured-performance} shows that the current implementation fits within the available system resources, and therefore further optimisation is not necessary.
We also note the value of conservative scaling estimates, to account for unexpected regressions during the implementation phase.

\subsection{Design}
\label{sec:design}

Having identified the top-level requirements of the system, we derived detailed requirements and identified possible suitable implementations.
While several options were considered, analysis showed that a highly integrated system where a single node type will handle all tasks, was the most attractive solution.
Each node will therefore need to
\begin{itemize}
\item receive data from LOFAR antenna fields
\item transpose this data
\item run a GPU correlator and/or beamformer
\item send the resulting data to the post-processing cluster
\end{itemize}

\subsection{Data flow}
\label{sec:dataflow}
\begin{figure}[htb]
\centering \includegraphics[width=.99\columnwidth]{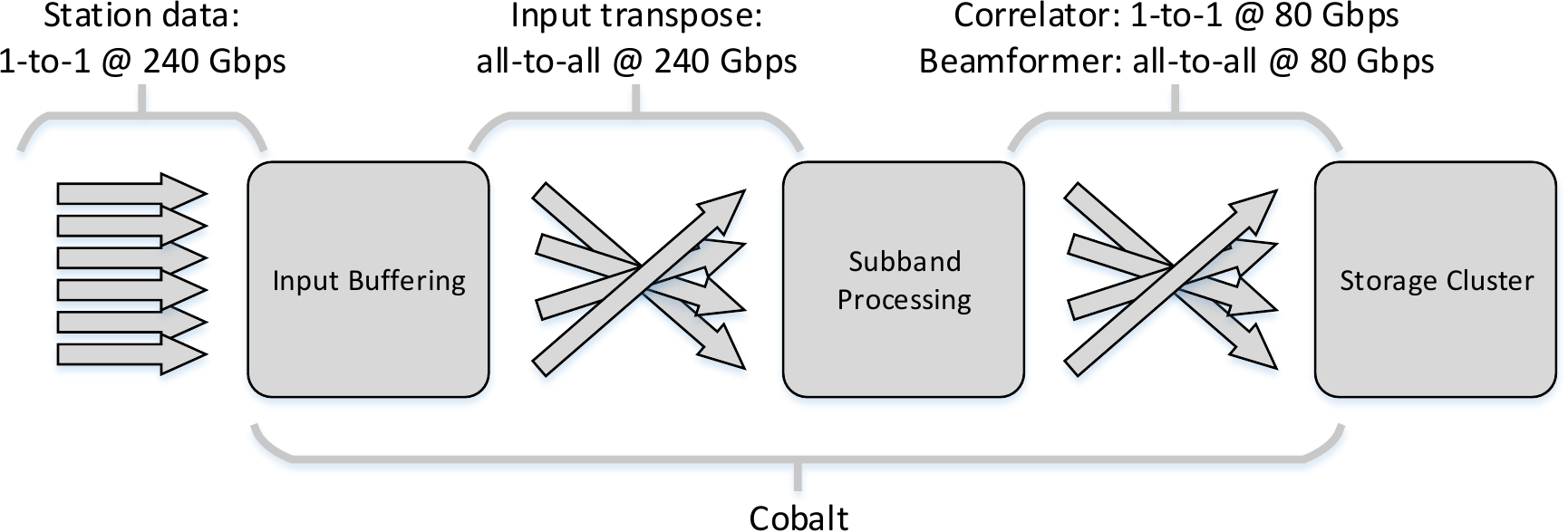}
\caption{Data flow from and to external systems, as well as within Cobalt.}
\label{fig:data-flow}
\end{figure}

A much simplified representation of the data flowing through the Cobalt system is presented in Figure \ref{fig:data-flow}.
Station data streams into the system at up to 240\,Gbps, using UDP/IP over Ethernet.
Output data is sent to the storage cluster, also using Ethernet, but here we are free to choose a reliable protocol such as TCP/IP instead.

Within the Cobalt system we need to fully reorder the data.
The data from antenna fields contains all frequency bands for a single antenna field, while the correlator requires data from all antenna fields for a single frequency band.

The LOFAR core network is based on 10 Gigabit Ethernet (GbE) technology.
Data from up to three LOFAR antenna fields is sent through each 10\,GbE link.
Combined with the required number of supported antenna fields, this gives a lower bound on the number of 10\,GbE ports we require in Cobalt (a minimum of 22, we designed for at least 27).
At this stage of the design process we considered 40 Gigabit Ethernet a viable and more dense alternative to four 10\,GbE ports.

The reordering of large volumes of data was considered a risk.
The efficiency of such a transpose, and the achievable bandwidth of the required low-latency interconnect, are difficult to estimate.
To mitigate this risk, we considerably over-dimensioned the network intended for this operation.
The transpose bandwidth is the same as the input bandwidth.
Our design target was to provide double the input Ethernet bandwidth specifically for the transpose.
Each Fourteen Data Rate (FDR) Infiniband Host Channel Adapter (HCA) provides a theoretical maximum achievable bandwidth of 54.54\,Gbps.
We therefore designed our system to provide one FDR Infiniband port for every two 10\,GbE ports, noting that this ratio needs to apply for every node.

\subsubsection{Memory bandwidth}
\label{sec:memory-bandwidth}
\begin{figure}[htb]
\centering \includegraphics[width=.85\columnwidth]{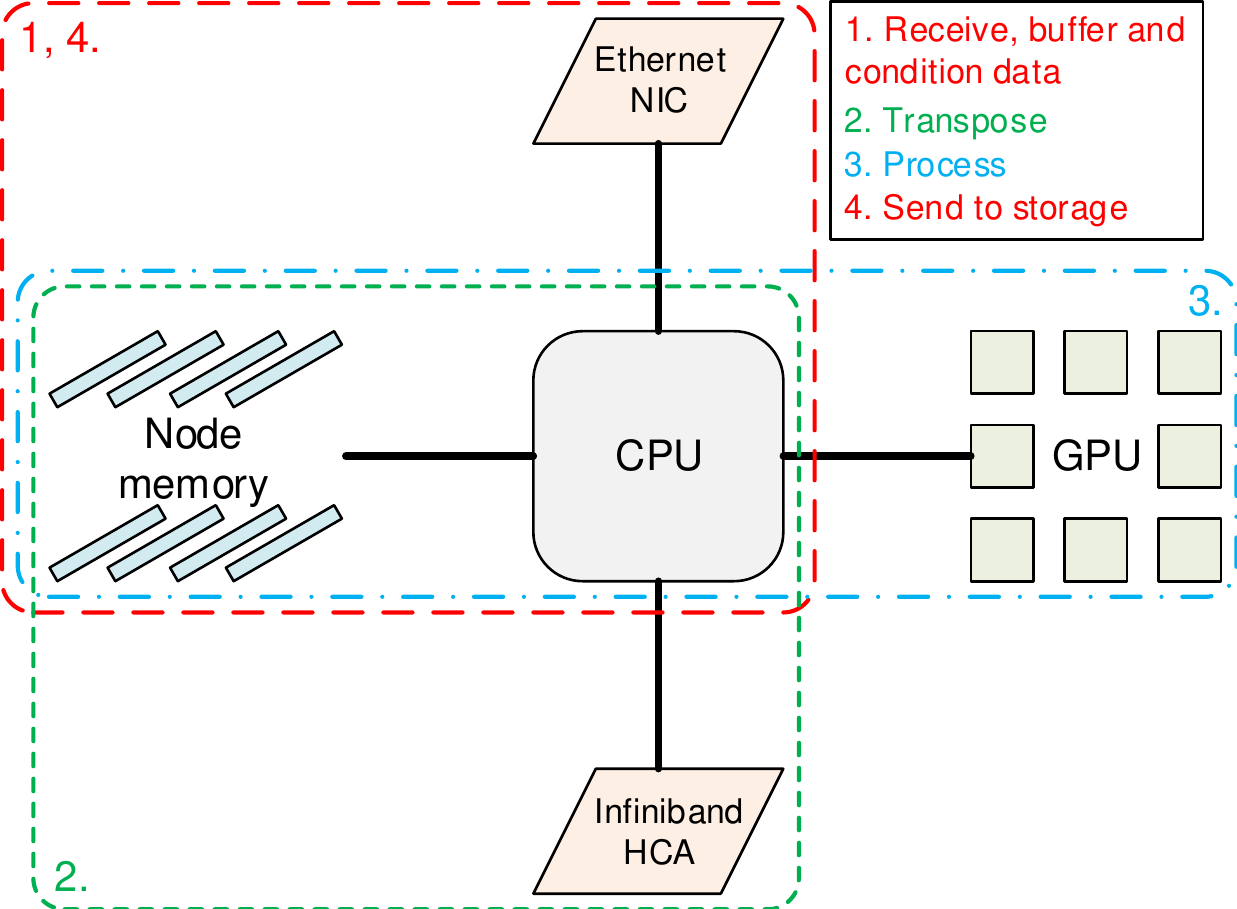}
\caption{Mapping of the various Cobalt tasks onto node hardware. This shows that node main memory, and in particular the memory  bus, is used for each task, highlighting a possible bottleneck.}
\label{fig:hw-mapping}
\end{figure}

The Cobalt system is characterized by a sustained and high rate of data streaming into the system.
This data stream needs to be received, conditioned and processed without loss.
Modern general-purpose operating systems are inherently inefficient at receiving data, due to the need to copy data several times before an application can access it\footnote{While this is an essential security feature, avoiding this potential bottleneck, for instance by the use of Remote Direct Memory Access (RDMA), is an active area of research.}.
This puts a considerable load on the memory subsystem, in particular on the available memory  bandwidth.
Figure \ref{fig:hw-mapping} shows the way the various tasks described in Section \ref{sec:design} were expected to be mapped on hardware.
We noted that the main memory bus was a potential bottleneck. 

An analysis of the memory bandwidth requirements was undertaken to estimate the system requirements in this respect, based on the input bandwidth and the number of times data is expected to be copied.
Handling of input was expected to drive this requirement, all other tasks combined were estimated to take less memory bandwidth.
The impact of hitting a memory bandwidth bottleneck was estimated to be high, we therefore took a conservative approach and limited maximum memory bandwidth use to 50\%.
Caching effects may positively affect used memory bandwidth, but are exceptionally unpredictable and were therefore not considered.
This, combined with the available memory bandwidth in the most recent Intel Xeon generation available at the time, gave us a lower bound on the minimum number of processors, and thus nodes, needed in the system.
Cobalt would require a minimum of six dual socket nodes in order to provide the required memory bandwidth, and our design target required eight dual socket nodes.

\subsubsection{Selecting the accelerator}
\label{sec:accelerator}



In selecting a suitable accelerator, we evaluated device specifications, performance of prototype code (per Watt and per Euro), software quality and programming environment.
Three vendors were evaluated: Nvidia, AMD and Intel, with a total of four devices investigated in more detail.

Although Intel's Xeon Phi was commercially available at the time, prototype code on this accelerator performed poorly due to the early state of the software stack.
It was therefore not considered further.
Both Nvidia and AMD had two devices available that would suit our applications.
The AMD FirePro S9000 and S10000, as well as the Nvidia Tesla K10 and K20X were evaluated in more detail, shown in Table \ref{tab:gpu-options}.

\begin{table*}[ht!]
  \centering
  \resizebox{\textwidth}{!} {
    \begin{tabular}{lcccc}
      \toprule
      & Nvidia Tesla K10 & Nvidia Tesla K20X & AMD FirePro S9000 & AMD FirePro S10000 \\ 
      \cmidrule(r){2-5}
      Architecture & Kepler       & Kepler         & Tahiti PRO    & Tahiti PRO   \\
      GPU          & 2x GK104     & 1x GK110       & Tahiti PRO GL & 2x Zaphod    \\
      Single Precision (GFLOPS)  & 4577         & 3935           & 3225.6        & 5913.6       \\
      Double Precision (GFLOPS)  & 190.7        & 1312           & 806.4         & 1478.4       \\
      Memory (MB)  & 2x 4096      & 6144           & 6144          & 2x 3072      \\
      PCIe         & PCIe 3.0 x16 & PCIe 2.0 x16   & PCIe 3.0 x16  & PCIe 3.0 x16 \\
      Programming  & Cuda         & Cuda           & OpenCL        & OpenCL       \\
      \bottomrule
    \end{tabular}
  }
  \caption{Considered GPU options.}
  \label{tab:gpu-options}
\end{table*}

Experience with prototype code showed that AMD devices generally performed better, but software and drivers stability for these devices was a potential problem.
This was considered unacceptable for a system that is an integral part of an operational instrument.
The Nvidia devices, although providing less computational performance, were superior in terms of stability, software quality and programming environment.
The Cobalt system does not require extensive double precision floating point support. 
The data-driven nature of the processing made support for PCIe v3 a secondary requirement, which K20X does not support.
Coupled with the superior single precision performance and lower energy consumption, Nvidia's Tesla K10 was selected as the accelerator of choice.
Cuda was selected over OpenCL as a programming model to take advantage of the superior debugging and profiling tools available, at the cost of having to rewrite the OpenCL based prototype code.
This selection, combined with the analysis in Section \ref{sec:requirements}, gives a lower bound on the number of accelerators required for Cobalt.
A minimum of 10 K10s (42.8 TFLOPS / 4.577 TFLOPS = 9.6) were needed.
Our design target required at least 14 K10s (61.3 TFLOPS / 4.577 TFLOPS = 13.4).

\subsection{Prototyping}
\label{sec:prototyping}

In Table \ref{tab:lower-bounds} we show a summary of the detailed lower bounds on the Cobalt system.
Based on the lower bounds discussed in the previous Sections, and a first order approximation of what may be a suitable node design, a list of components for Cobalt was proposed.
\begin{table}
  \centering
  \resizebox{\columnwidth}{!} {
    \begin{tabular}{lccc}
      \toprule
      & Minimum & Design target & Proposed Cobalt\\
      \cmidrule(r){2-4}
      Nodes & 6 & 8 & 8\\
      10\,GbE ports & 22 & 27 & 32 \\
      FDR HCAs & 11 & 14 & 16\\
      Nvidia Tesla K10s & 10 & 14 & 16\\
      \bottomrule
    \end{tabular}
  }
  \caption{Detailed lower bounds for the Cobalt system.}
  \label{tab:lower-bounds}
\end{table}

Based on the lower bounds identified above, we proposed a baseline Cobalt system that consisted of at least 8 nodes.
Each of these nodes would have four 10\,GbE ports (or equivalent), two FDR Infiniband ports and two accelerators.
We noted that dual-port FDR Infiniband HCAs are inherently bottlenecked by their limited PCI-express bandwidth, so two single-port HCAs were required.
Our task next task was to find a suitable commercially available node, and evaluate a representative sample for performance.
The entire product line of all major vendors was evaluated, based on suitability, availability and maintainability..
Having surveyed a large number of nodes from a variety of vendors, we selected our initial prototype based on a Dell R720 chassis.
This node had a single 40\,GbE port instead of the four 10\,GbE ports, but matches all other requirements.


\subsubsection{PCI-express balancing}
\begin{figure*}[hbt]
  \centering
  \begin{subfigure}[b]{0.35\textwidth}
        \includegraphics[width=\textwidth]{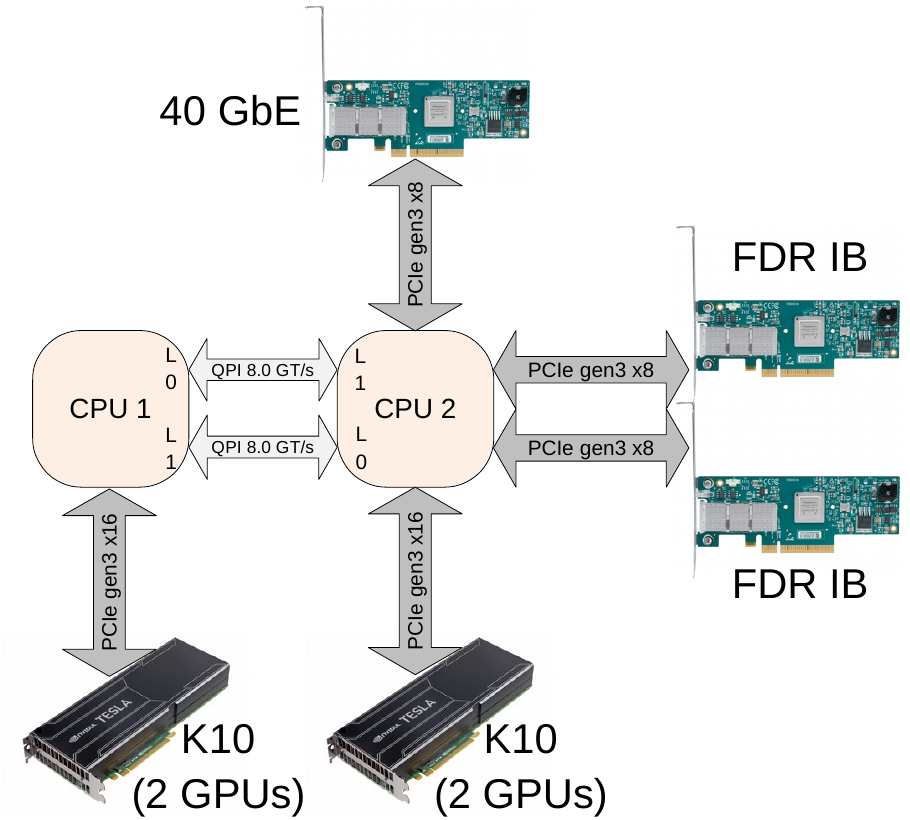}
        \caption{Dell R720}
        \label{fig:R720-PCIe}
  \end{subfigure}
  ~ 
  \begin{subfigure}[b]{0.50\textwidth}
    \includegraphics[width=\textwidth]{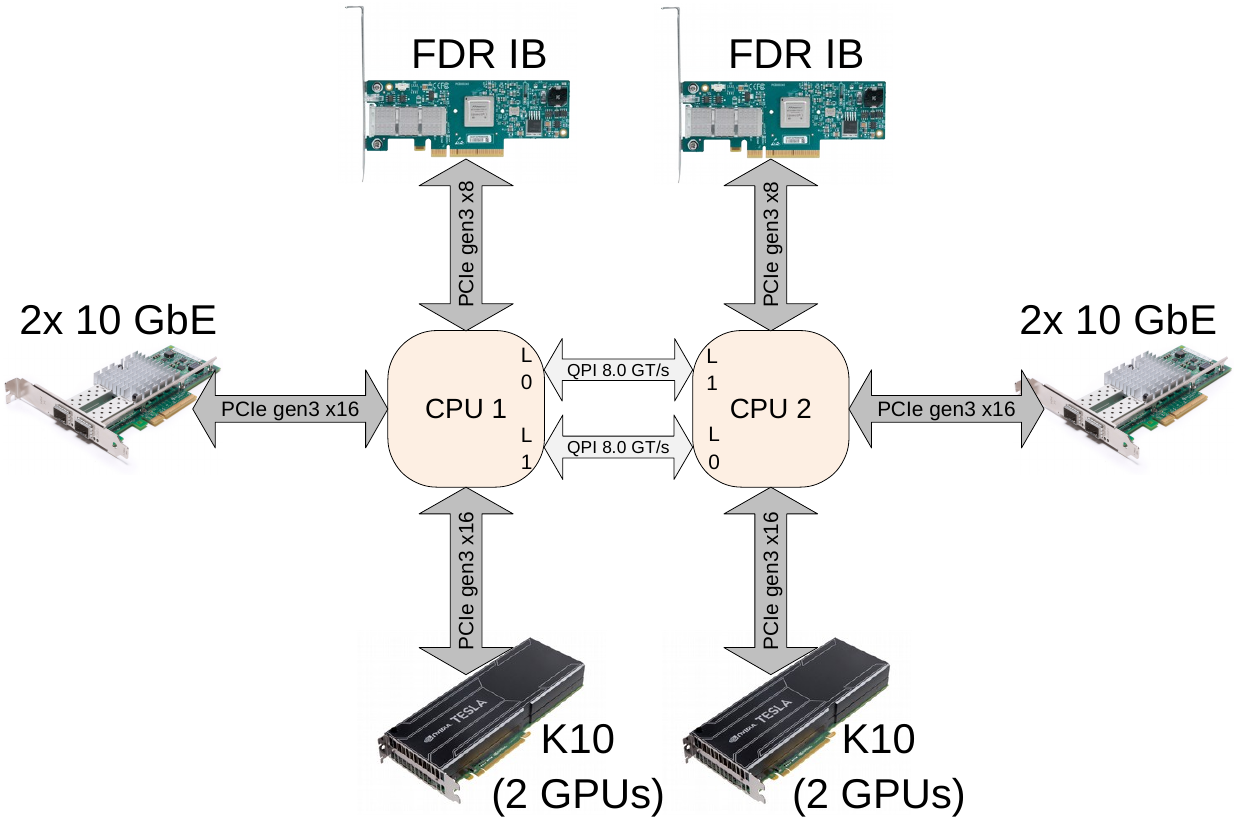}
    \caption{Dell T620}
    \label{fig:T620-PCIe}
  \end{subfigure}
  \caption{PCIe configurations encountered in the two prototype systems.}
\end{figure*}

The primary data transport interfaces in Cobalt nodes is PCI-express (PCIe).
Our system consists of many inter-communicating components, so a well balanced PCI-express infrastructure is vital to an efficiently operating correlator and beamformer.
We investigated the configuration of a prototype Cobalt node, the standard Dell HPC node at the time, a Dell R720 (shown in Figure \ref{fig:R720-PCIe}).
In this figure, a clear imbalance can be seen, as the vast majority of PCIe connectivity is provided by a single CPU.
All data for the other CPU, or the accelerator attached to that CPU, had to cross the Quick Path Interface (QPI) boundary between CPUs at least twice.
Based on experimental data, it was considered highly likely that this would be a significant bottleneck.

Finding a system that exposes next to all the available PCIe lanes in a more balanced manner, turned out to be quite difficult.
Figure \ref{fig:T620-PCIe} shows the configuration of a Dell T620 workstation node.
Even though these nodes were not specifically designed for HPC use, the balanced PCIe configuration shown led to this being selected as our base node type.
These nodes also allowed for the installation of two dual-port 10\,GbE Network Interface Controllers (NICs), in place of the single port 40\,GbE NIC in the R720 that was found to be unsuitable in the existing 10\,GbE network.
In Section \ref{sec:software-architecture} we leverage the symmetrical architecture of these nodes by essentially using them as two mostly independent nodes, one for each CPU socket, both for clarity and performance.

\subsubsection{Cooling the GPUs}
\begin{figure}[htb]
  \centering
  \begin{subfigure}[b]{0.4\textwidth}
    \includegraphics[width=.8\columnwidth]{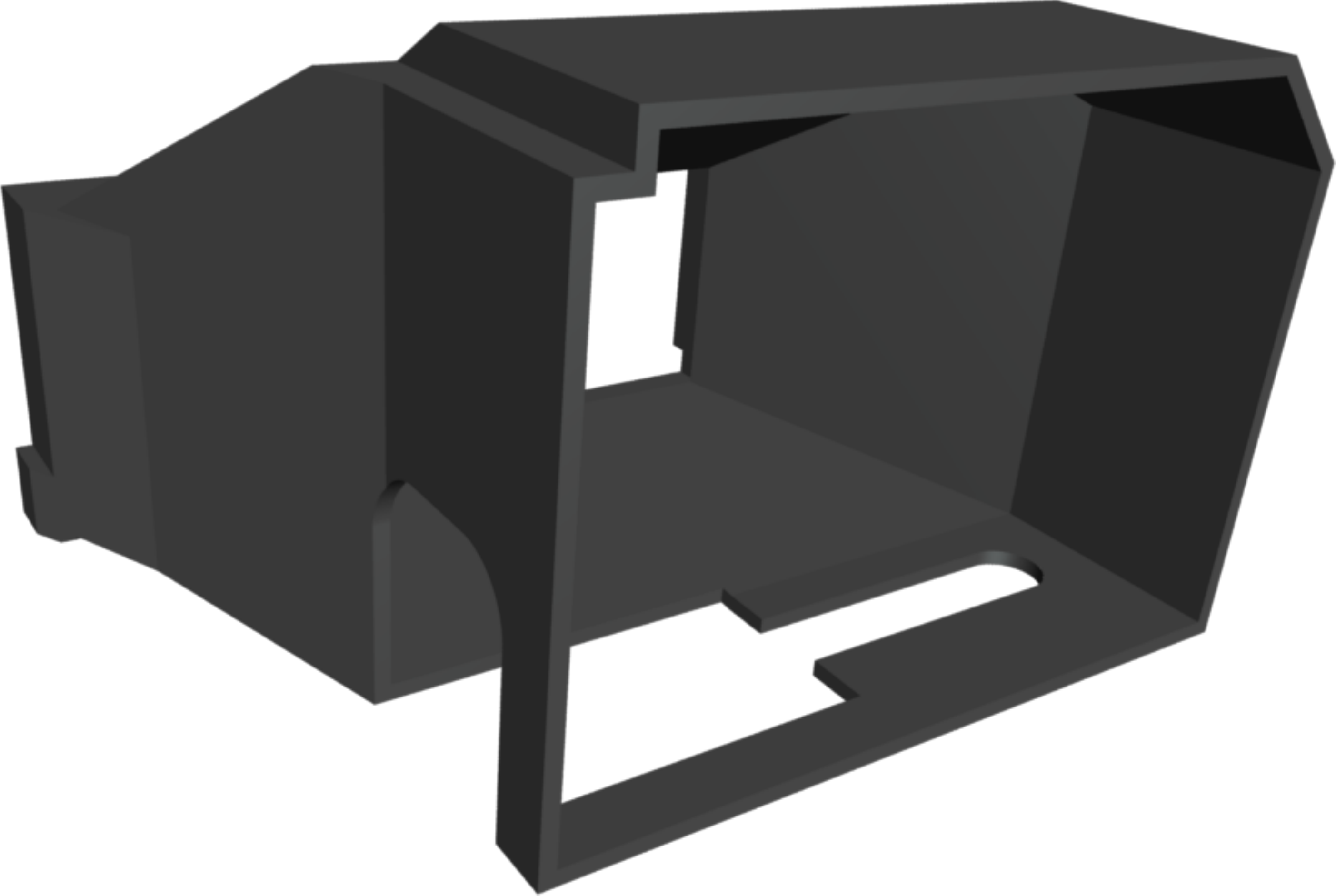}
    \label{fig:Cobalt-baffle-design}
  \end{subfigure}
  ~
  \begin{subfigure}[b]{0.4\textwidth}
    \includegraphics[width=.8\columnwidth]{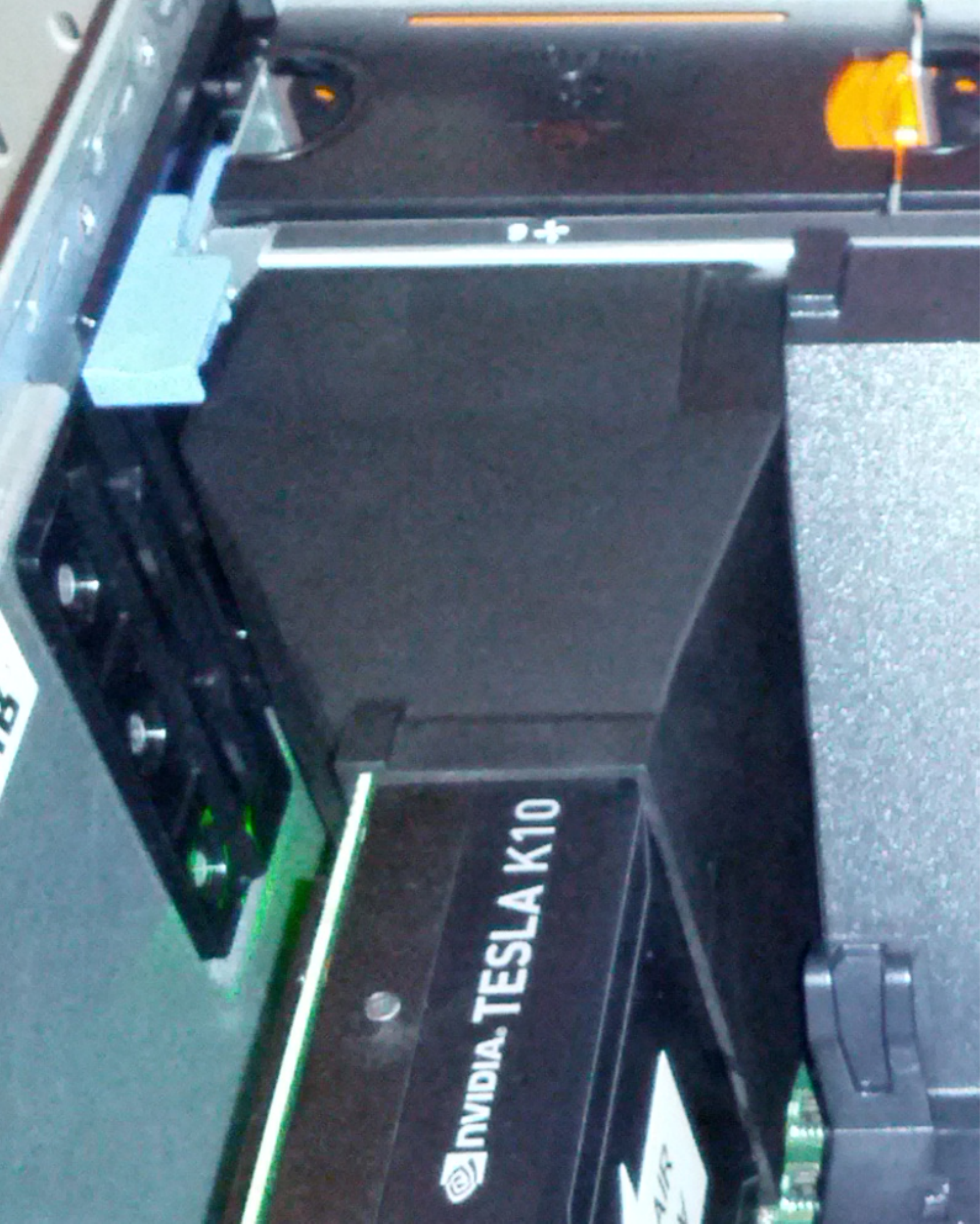}
    \label{fig:Cobalt-baffle-installed}
  \end{subfigure}
  \caption{3D render of the production Cobalt airflow baffle (top) and a Cobalt node with custom baffle installed (bottom). Note the fan at the top of the image, providing forced air cooling to the Nvidia Tesla K10 via the installed custom duct.}
  \label{fig:baffle}
\end{figure}

The Dell T620 chassis was designed as a workstation, rather than a HPC node.
The Nvidia Tesla K10 was only available as a passively cooled unit, which relies on the chassis to provide sufficient cooling.
These two facts combined meant that we ran into serious cooling issues for the selected GPUs.
Early tests showed that the K10s ran at approximately 70$^{\circ}$C while idle, with an optional fan-bar installed.
No load tests could be performed, since the GPUs would overheat and switch off before any meaningful test results could be obtained.
Improvised cardboard and ductape airflow baffles showed that sufficient cooling could be provided to the GPUs.
Better fitting baffles were designed and 3D printed in-house at ASTRON.
By directing the airflow generated by the fan bar through the Nvidia Tesla K10 GPUs, we successfully reduce the operating temperature of the GPUs to acceptable levels.
Using these custom baffles, shown in Figure \ref{fig:baffle}, the Dell T620 and Nvidia Tesla K10 combination ran about 10$^{\circ}$C cooler than a comparable Dell R720 system, probably due to the additional space in the (large) Dell T620 chassis.
We outsourced the production of twenty of these baffles by injection molding rather than 3D printing, sufficient for ten Cobalt nodes.

\subsection{The Cobalt system}
\label{sec:cobalt-hw}
Apart from the issues described above, no other performance limitations were identified with the Dell T620 nodes.
The fully deployed Cobalt system consists of ten of these nodes, eight production and two hot spare and development nodes, fitted with two Nvidia Tesla K10 GPUs each.
Each node contains two dual-port Intel X520 10\,GbE NICs and two single-port Mellanox ConnectX-3 FDR HCAs.

\section{Software design}
\label{sec:software}

The software part of Cobalt consists of two applications that manage the data flow through networks and GPUs, and store correlated and/or beamformed data products on persistent storage as shown in Figure~\ref{fig:data-flow}.
Cobalt also interfaces with several other subsystems for control, monitoring, logging, and metadata.
No data is fed back from post-processing into Cobalt.



The following Subsections describe the Cobalt software architecture and considerations for parallelism at different layers.

\subsection{Software architecture}
\label{sec:software-architecture}

\begin{figure*}[htb]
  \centering \includegraphics[width=0.8\textwidth]{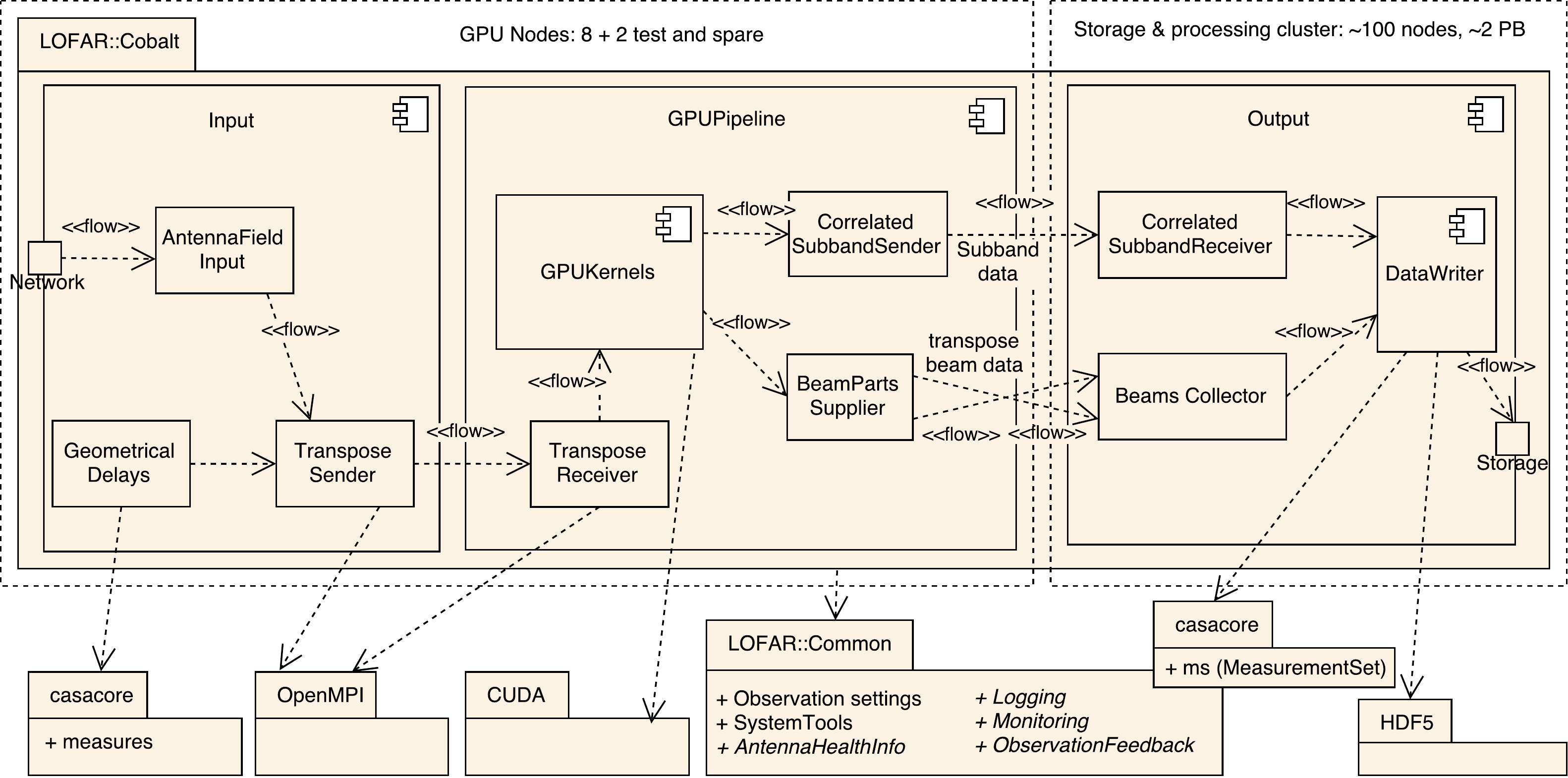}  
\caption{High-level component diagram of the LOFAR Cobalt software with data flow and dependencies.}
\label{fig:software-components}
\end{figure*}

The component diagram in Figure~\ref{fig:software-components} shows high-level LOFAR Cobalt components (here named in \texttt{typewriter} font), dependencies and data flow.

Observation control starts \texttt{Output} processes on all allocated nodes in the storage cluster.
It then uses MPI (Message Passing Interface) to start two processing applications (MPI ranks) per GPU cluster node, one per CPU socket.
Each data processing instance connects to \texttt{Output} processes it needs to send data to, and opens sockets for its two 10\,GbE interfaces to receive antenna field data.
Just after the observation start time, data flows through Cobalt producing data products on the storage cluster.
On late establishment or failure of network connections, Cobalt retries until the observation stop time.
Then, observation meta data such as LOFAR system health statistics are gathered from databases and written into the data products.
Before shutting down, Cobalt gives its vote for the observation end status to LOFAR control.

All software components along the data flow path forward data blocks of about 1 second using MPI, thread-safe bounded FIFO queues or TCP/IP.
We allocate block space once during initialization and keep pools of free blocks.
The block size is a trade-off: efficient network transfers and processing favor larger blocks, but the size is limited by GPU memory (4 GiB), and affects how many beams the beamformer can form, as well as main memory footprint and overall latency.
The exact block size is a multiple of all work unit sizes in each signal processing step to limit the number of edge cases to implement, test and debug.

Each \texttt{AntennaFieldInput} receives UDP datagrams on two 10\,GbE network interfaces and forwards valid data to the \texttt{TransposeSender}.
\texttt{Transpose} uses a circular buffer to perform coarse delay compensation by shifting the sample streams by an integer number of samples ($\sim$5.12 \textmu s).
These delays are computed by \texttt{GeometricalDelays} in a separate CPU thread, and are used to compensate for different signal arrival times at different antenna fields and to form beams.
The remaining (sub-sample) delay is compensated for later using phase rotation on the GPU.

\texttt{TransposeSender} also deals with (rare) out-of-order UDP datagrams and drops data that arrives after a deadline.
\texttt{TransposeReceiver} in the \texttt{GPUPipeline} component transposes data per antenna field to data per subband using MPI over Infiniband.
The \texttt{GPUPipeline} component pushes subband data through the signal processing pipeline on GPUs, producing correlated and/or beam data, as explained later.
Each correlated subband is sent to an \texttt{Output} component on a single host using TCP/IP, but beam data needs to be transposed over the network to collect all subbands for each beam, produced at different GPUs, to be combined in a single storage host.
The \texttt{Output} component stores correlated data in the MeasurementSet format using casacore\footnote{\url{https://github.com/casacore/casacore}} and beam data in the LOFAR HDF5 format\footnote{\url{https://www.hdfgroup.org/HDF5/}}.

\subsection{Dealing with jitter and hardware failure}

Cobalt is part of a large, operational system and as such uses the LOFAR \texttt{Common} library to communicate with several monitoring and control systems, and to reuse other common functionality.
Antenna fields send data at a fixed rate, but contention on computing and especially on network and storage resources may vary.
As a complex system with different sites, jitter, hardware failures and misconfigurations do sometimes occur.
We therefore designed Cobalt to conceivably drop data rather than fail or wait in several key places.
The network or operating system may drop incoming UDP data, the \texttt{TransposeSender}'s circular buffer may drop data if not read out in time.
Both \texttt{CorrelatedSubbandSender} and \texttt{BeamPartsSupplier} have bounded queues that drop when full.
Any overload or failure in the pipeline will fill the previous component's queue, propagating until such a dropping point is reached.
Cobalt encodes lost or dropped data in metadata that is aggregated and written into the data product for post-Cobalt processing to interpret.

We routinely correlate 488 subbands (about 96\,MHz wide) from up to 78 antenna fields (230\,Gbps input) or produce 222 beams (37\,Gbps) from 12 antenna fields (or a compromise of both) using 80 storage and 8 GPU nodes.
Correlation is GPU compute-bound, but for beamforming output bandwidth to the storage cluster is the limiting factor, which is not bound by Cobalt.
Most beamforming science needs high time resolution and as many beams as we can form, up to the available capacity.
Measurements show that up we can form up to 146 beams for 288 16-bit subbands, which is well in excess of our original requirement, even if we cannot store the resulting beams at the desired time resolution.

\subsection{Workload distribution}

The Cobalt hardware is fundamentally different from its predecessor, the IBM Blue Gene/P supercomputer.
In Blue Gene/P we needed several cores to process a single subband, but in Cobalt a single GPU is powerful enough to process several subbands.
In Blue Gene we designed a complex round-robin work-distribution scheme to avoid contention on the internal torus network~\citep{Romein:10}.
In Cobalt a static assignment of subbands to GPUs is sufficient.
Table \ref{tab:hardware-parallelism} shows the levels of hardware parallelism in Cobalt.

Table~\ref{tab:software-parallelism} indicates application data dimensions that we must map to hardware parallelism.
The independence within dimensions (e.g.\ process two antenna fields independently) is not available throughout the complete processing pipeline: at several points data has to be combined or forwarded jointly (i.e.\ synchronized).
In terms of scaling direction, most dimensions scale \emph{out}.
When adding more antenna fields, beamforming and correlation output also scale \emph{up}, the latter quadratically.

Apart from data parallelism, processing and I/O task parallelism are also possible: receive input data, geometrical delay computation, input data transposition, control of GPU data transfers and kernels, data transfer to storage, and data product write-back all run in parallel on the same hardware.
We leverage all levels of parallelism mentioned in Tables \ref{tab:hardware-parallelism} and \ref{tab:software-parallelism} to ensure we can keep up with the most demanding observation setups.

\begin{table}[hbt]
\centering
  \resizebox{\columnwidth}{!} {
    \begin{tabular}{lll}
      \toprule
      Layer     & Type (Qty.) & API \\
      \cmidrule(r){2-3}
      Cluster   & Multi-node (8), multi-CPU (16) & MPI \\
      Half-node & Multi-core (16 SMT), multi-GPU (2) & OpenMP, pthreads \\
      GPU       & SMs (16), cores (1536) & CUDA \\
      \bottomrule
    \end{tabular}
  }
\caption{Hardware available for parallel execution.}
\label{tab:hardware-parallelism}
\end{table}

\begin{table}[hbt]
  \centering
  \resizebox{\columnwidth}{!} {
    \begin{tabular}{ll}
      \toprule
      Data dimension   & Size (typical) \\
      \cmidrule(r){2-2}
      UDP datagrams    & 48828 per antenna field per second \\  
      Antenna fields   & 38--78 (correlator), 12--48 (beamformer) \\
      Freq.\ subbands  & 200--488 \\
      Freq.\ channels  & 64--256 (correlator), 1--16 (beamformer) \\  
      Samples (time)   & 768--196608 per freq.\ channel per second \\  
      Beams            & 1--222 \\  
      \bottomrule
    \end{tabular}
  }
\caption{Dimensions to map to (data parallel) hardware.}
\label{tab:software-parallelism}
\end{table}

We partition the antenna field streams over all 10\,GbE interfaces and the subbands over all MPI ranks and their GPUs.
Work partitioning and mapping to GPU resources is compute kernel specific.
To utilize all compute resources, a GPU needs to be supplied with many blocks each with many (semi-)independent work units.
I/O and memory access need to be carefully considered too, as many of our compute kernels are bound by GPU memory bandwidth.
Exact partitioning and mapping differs between observation setups, especially for dimensions that are traded off against each other.
For example, fewer frequency channels implies more samples in time, providing a different dimension for parallelism.
We compile our CUDA kernels at runtime to turn observation-specific constants into compile-time constants.
Run-time compilation increases performance by removing branches and by lowering the register pressure, and allows more freedom with respect to workload distribution within the GPU.
To control and process on GPUs we use CUDA~\citep{Nickolls:2008} and the CUFFT library.
In contrast, the Blue Gene/P PowerPC CPUs required handcrafted assembly to fully exploit their processing power.

\subsection{Parallelization libraries}

Cobalt uses \emph{OpenMP}, \emph{OpenMPI}, \emph{CUDA}, \emph{fork/wait} (for runtime kernel compilation), \emph{pthreads}, and \emph{signals} (to initiate shutdown) in the same processing application.
Some of these were not designed to work together and require careful programming.

To exploit task parallelism we need to determine task granularity and mapping, such that tasks both run and forward data blocks in time.
The \emph{OpenMP} pragma \texttt{omp parallel for} is an easy way to iterate over the subbands in parallel.
Around that we placed \texttt{omp parallel sections} to divide pipeline work into parallel tasks.
Tasks forward blocks through \emph{thread-safe bounded queues} that use \emph{pthreads} condition variables, not available in \emph{OpenMP}, to avoid busy waiting.
Although \emph{OpenMP} and \emph{pthreads} are not intended to be used together, this results in excellent readability of the multi-threaded code, while allowing the use of powerful primitives like \emph{thread-safe bounded queues}.
Multi-threading remains in a local scope and both data flow and control flow remain very clear.
Another upside is that our \emph{OpenMP} pipeline allows us to trivially adjust task granularity and count, without requiring a direct mapping to CPU cores.
Downsides include the non-portability of our combined use of \emph{OpenMP} and \emph{pthreads} and that this use favors to have as many threads as tasks, as otherwise some tasks have no dedicated thread and thus may not empty their input queue, causing deadlock.
As a result, some observation setups end up with an order of magnitude more threads than (logical) CPU cores.
While there is room for CPU task management, the current \emph{OpenMP} code is well readable and further optimizations will not improve \emph{system} capability, since CPU power has never turned out to be a bottleneck in our system.

The Infiniband and GPU cards need the same CPU memory used for DMA (Direct Memory Access) to be pinned and registered with their driver.
Pinning and registering come with an overhead, which we have mitigated by allocating all of these buffer during initialization.
Both the MPI and GPU library offer interfaces to explicitly allocate memory for DMA, but only CUDA can mark an existing allocation as such, so we allocate shared buffers via MPI and then register them with CUDA.

Cobalt deals with a lot of mostly independent data streams that are handled in parallel without interdependencies.
To optimally utilize the available hardware, every level of available parallelism needs to be exploited.
However, none of the MPI libraries we looked into offered good multi-threading support, they were either not thread-safe, used a global lock, or failed to compile or run with fine grained thread synchronization.
We therefore wrap our MPI calls with a global lock, which turns out to be efficient enough in combination with non-blocking sends and receives using \texttt{MPI\_Isend} and \texttt{MPI\_Irecv}.
We do need a separate polling thread to frequently check for completion of pending transfers using \texttt{MPI\_Testsome}, otherwise MPI throughput suffers.

On the storage cluster, we distribute all subbands and beams over the nodes.
Some beamforming observations need full resolution, both spectral as well as temporal, which limits the number of beams that can be sent to storage due to limited network bandwidth.
In such setups, we have to store each beam across multiple storage nodes.
This split is less convenient for post-Cobalt processing, to be executed on the same cluster.

\subsection{Signal processing with GPU kernels}
\label{sec:gpu_kernels}
This Subsection focuses on the digital signal processing GPU kernels shown in Figure~\ref{fig:signal-processing} as executed within the \texttt{GPUKernels} component.


The correlator pipeline first channelizes subbands in a polyphase filter using FIR filters and FFT kernels.
We carry FIR filter history samples across to the next block.
The pipeline then applies fine delay compensation and bandpass correction.
This marks the end of processing per antenna field.
To efficiently operate \emph{across} antenna fields, the delay and bandpass kernel transposes data on write-back to GPU device memory.
The last kernel computes the correlations of all pairs of antenna fields and averages in time to approximately 1\,s.

The beamformer pipeline forms many \emph{coherent} and/or \emph{incoherent} beam(s).
Both beam types have the first four kernels in common.
Cobalt performs delay compensation, bandpass correction and beamforming at 256 channels per subband as a good compromise between time and frequency resolution, and then transforms to the requested output resolution, often 1 or 16 channels per subband.
After bandpass correction, the coherent and/or incoherent specific steps of the beamforming pipeline execute.
Coherent beamforming first adjusts the beam direction with a phase shift and sums over antenna fields, then optionally computes Stokes parameters, while incoherent beamforming first computes Stokes parameters and then sums over antenna fields.
Coherently formed beams are more sensitive but cover a much smaller sky area.
During an observation many adjacent beams can be formed to mosaic a somewhat larger sky area, although some projects also add an incoherent beam to quickly search for bright signals~\citep{coenen:2014}.
If we do not convert to coherent Stokes I (intensity only) or IQUV (full polarization), we retain complex voltage data with phase information allowing coherent dedispersion (after Cobalt).
However, complex voltages cannot be time averaged.

Due to differences in required frequency/time resolution and averaging, the beamformer and correlator pipelines diverge quickly in how they transform the incoming signal.
This means that our beamformer cannot share initial steps with the correlator and needs to reorder data often as shown in Figure~\ref{fig:signal-processing}.

All kernels operate on single-precision complex floating-point data, except for delay compensation, which uses mixed precision.
Fine delay compensation (i.e.\ subsample) uses the residual delay from coarse delay compensation by the \texttt{TransposeSender}.
From the residual delays at the start and end of a block, we compute the channel-dependent phase angles in double precision.
Within a 1\,s block these angles can be interpolated linearly to obtain the angle for each sample.
Then back in single precision, we determine the phase shift factor (sin/cos) and rotate back the phase of each sample (complex multiplication).
The beamforming kernel operates in a similar way to form beams with an offset from the center.
The Tesla K10 GPU has low double precision throughput, but as long as the kernel is memory bound, the limited use of double precision has little impact.

Most kernel parameters are fixed throughout an observation.
We avoid using registers for these parameters and obtain more efficient kernel binaries by using runtime compilation supplying fixed parameters as C-style defines.
The resulting code is also more readable.
We reduce GPU memory usage by using a small number of buffers that the CUDA kernels alternate between as their in- and output.


The number of observation parameters supported by Cobalt is large.
This affects kernel complexity, kernel execution configuration (CUDA block and grid dimensions), as well as input/output data dimensions and some transpose alternatives.
This complexity cannot lead to observation failures.
To deal with execution configuration, GPU buffer sizes and performance counters, we use a wrapper class for each kernel.
This also wraps the type unsafe argument passing when launching a CUDA kernel. 
Each kernel unit test covers the wrapped kernel.
Furthermore, we centrally document which buffers are (re)used by which kernels and what the array dimension order and sizes are.

Although the development of highly optimized GPU kernels is a critical Cobalt ingredient, the details are outside the scope of this article.
For more insight into radio astronomy signal processing for Cobalt and beyond on various accelerator platforms, we refer the interested reader elsewhere~\citep{romein:16}.

\section{Verification and validation}
\label{sec:verification}

Before Cobalt could be taken into operational use it needed to be extensively tested and tuned.
Regression testing and integration happened continuously during (software) development.
We determined science readiness during \emph{commissioning}, a phase in the last part of development where domain experts and instrument engineers work closely together towards system-wide integration, validation, tuning and performance characterization.
Some of these tests are still performed on one Cobalt node and LOFAR station before deploying a new software release at full scale.


During Cobalt development we added about 400 tests in 100 test programs.
Some are unit tests, others test a feature, uncommon observation settings, across an interface, or a complete Cobalt pipeline on a tiny amount of data.
About another 100 unit tests were already in place for the \texttt{LOFAR Common} package. 

Incrementally developing tests was a substantial amount of work.
Extending the test set and updating documentation are part of delivering a new feature.
What added to the effort was dealing with tests that generally pass, but occasionally fail due to race conditions or non-real-time testing of real-time code.
We used the Jenkins\footnote{\url{https://jenkins.io/}} \emph{continuous integration} service to manage regression test builds.
The extensive use of testing was critical for Cobalt to minimize regressions, both on component and on system level.
Furthermore, tests kept the code maintainable, by providing confidence and freedom to improve or even refactor the Cobalt code.

Cobalt needs various non-default system settings to perform well.
System firmware (BIOS/EFI) and Linux kernel settings needed to be tuned for performance and predictability, such as (minimum) network buffer sizes, CPU frequency scaling, and mapping GPU and NIC interrupts to the CPU they are linked to.
We do not need to bind threads to cores within a socket, as long as we raise the CPU and I/O priority of threads receiving UDP input and writing to storage.
We also do not need to run a PREEMPT\_RT (real-time) patched Linux kernel.
Our multi-homed network and VLANs to international stations required changes to ARP and routing settings to function correctly.

To get good performance for the input transpose via MPI, we needed to tune OpenMPI RDMA settings, for which we used the point-to-point tests from the SKaMPI benchmark~\citep{Reussner:2002}.
We also send transfers between CPU sockets over infiniband instead of directly between the CPUs via the on-board QuickPath Interconnect.

Due to a performance regression that couldn't be resolved by reverting code commits, we had to rework the MPI transfer scheme.
Instead of supporting all surrounding tasks independently by scheduling their many point-to-point transfers, we applied message combining to send fewer but larger messages.
While the new implementation solved the performance regression, this came at the cost of increased use of memory/cache bandwidth, and it introduced more dependencies between producers and consumers of MPI data.
This is an example where we sacrificed an over-dimensioned resource (CPU cache/memory bandwidth) for a scarce resource (development effort).


We have more examples of unexpected regressions during development and operations, but in general, debugging performance issues silently introduced with system software updates, changed system \& network settings, or replaced hardware was time consuming and difficult.
To alleviate this risk, we used performance and configuration verification scripts.
This \emph{operational readiness check} was especially useful when the line between responsibilities for high performance software and system and network administration blurred.
When major hardware/software functionality had passed verification, the project scientist (i.e.\ an Observatory astronomer) was responsible for the validation effort to deliver a science capable instrument.




Radio telescopes essentially sample electromagnetic noise, including radio interference, and then perform stochastic signal processing.
Thus there was no reference output to bit-wise compare our output to.
Moreover, the existing BlueGene-based system used double precision and a different beamformer DSP filter chain.
We therefore chose to analyze Cobalt output to comply with signal and noise properties required for the most demanding science cases.
This proved the validity of the Cobalt output without having to be bit-wise equal to its Blue Gene predecessor.

In total, we planned and performed about 30 experiments and worked with astronomers and software developers to get issues resolved and the system tuned and characterized.
This effort took several months.
Several experiments required custom tools or software hooks and resolving issues can be time consuming.
This was a substantial project risk that had to be mitigated with a solid development process and extensive and early testing.

During commissioning we observed no perceptible increase in system noise between the Blue Gene/P based correlator and beamformer and the new Cobalt implementation.
Considering the difference in numerical precision used -- double precision in Blue Gene, single precision in Cobalt -- this warrants some discussion.
We note that these choices were driven primarily by the selected hardware, not by necessity.
Blue Gene was designed for double precision processing. There was no advantage in using lower precision arithmetic.
In contrast, the selected Nvidia K10 GPU is optimised for single precision processing.
As shown in Table \ref{tab:gpu-options}, this GPU has abysmal double precision performance.
Only delay compensation was considered vulnerable to this loss of precision.
Comparative analysis showed that single precision delay compensation led to an insignificant increase of the total noise~\citep{CobaltCommissioning}.
Calculating the delays themselves does require double precision, this is the only part of the Cobalt pipeline to do so.

\section{Operational experience}
\label{sec:operational}

Cobalt has been LOFAR's secondary correlator since January 2014 and its primary since March 2014.
In May 2014, Cobalt also took over for beamformer observations.

\begin{figure}[hbt]
  \includegraphics[width=\columnwidth]{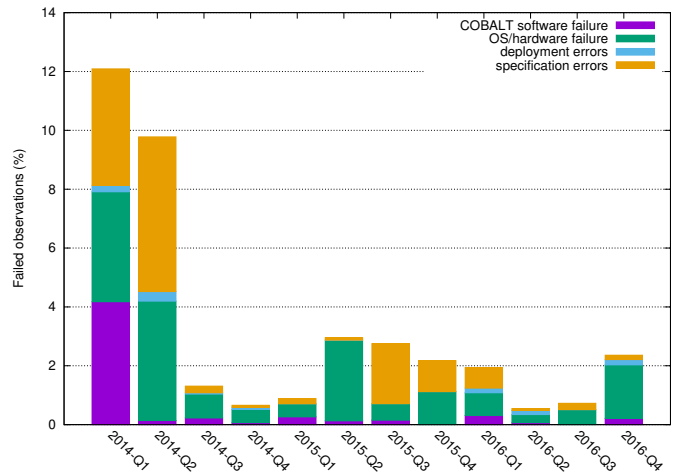}
  \caption{Failure modes of the Cobalt system over three years, and their occurrence in percentages.}
  \label{fig:failure_statistics}
\end{figure}

We have collected statistics from three years of operations with the Cobalt system.
Figure \ref{fig:failure_statistics} shows the relative number of failed observations, with a break down into four failure modes ($N \approx 23000$).
On average, 97.3\% of submitted observations were successful, clearly exceeding the required operational availability of $>95\%$ described in Section \ref{sec:system-requirements}.


Observations may start at any moment (24/7), but normally, issue investigation starts the next working day.
Observations scheduled between the occurrence of an issue and the start of the next working day may be adversely affected.
The availabilities of other LOFAR sub-systems are not shown, but were generally lower than that of Cobalt.
However, these generally work on non-volatile data where failures do not automatically result in irretrievable data loss.

There was a noteworthy increase in availability after six months in operations that can be attributed both to burn-in, as well as bug fixes in the Cobalt software and in the scheduling system.
Current operational Cobalt failures mostly originate in network configuration or services, or in non-standard observation settings.
Hardening and monitoring the network and settings have reduced their impact (until such monitoring services fail).
We have run into several Linux kernel bugs, unexpectedly exposed with new software releases or changes in work load.
This includes a failure mode that caused occasional Linux kernel panics in our system, due to a memory allocation bug that was fixed with a newer kernel release.
While this shows the value of keeping low-level software updated and patched, we note that the regression mentioned in Section \ref{sec:verification} may in part have been caused by similar updates.



\section{Summary and discussion}
In this paper we presented the Cobalt GPU-based correlator and beamformer system for the LOFAR radio telescope.
This system has successfully replaced the earlier Blue Gene based systems and has been in operations for almost four years now.
We introduced the hardware design, as well as the data flow-driven simplified system engineering process that led to the final implementation. 
The challenges that were faced during prototyping were described, as well as some of the engineering efforts that were necessary to keep the GPUs at an acceptable operating temperature.
Finally, we showed some of the details of the software design, the verification process, and we discussed the operational experience with the Cobalt system.

All nodes in Cobalt are identical and perform all necessary processing, there are no dedicated nodes for a task.
This requires careful programming, as was shown in Sections \ref{sec:software} and \ref{sec:verification}, but also makes for a highly efficient system with few idle components.

In contrast to similar papers describing software correlators, we focused heavily on the development process of the system design.
We showed how hardware/software co-design, in close collaboration with a commercial partner, can lead to an efficient and affordable system.
None of the systems aimed at the HPC market were, for various reasons, suitable for our application.
Close interaction between hardware vendor, hardware system designer and software architect in the design and prototyping phases was instrumental in finding a suitable node design.

An Agile test-driven development process was introduced to ensure timely delivery of a system that is fit for purpose and meets the requirements described in Section \ref{sec:system-requirements}.
We also noted in Section \ref{sec:verification} that the test-driven aspect had great advantages in a system that cannot be completely deterministic.
As another example of co-design, the experiences with previous LOFAR beamformer systems showed that a redesign of this component would better match the requirements of the majority of the science users.
While this delayed the delivery of the Cobalt beamformer slightly, we took this opportunity to improve LOFAR non-imaging capability.



\section{Impact}
This project has generated a surprising amount of interest.
Discussions with the University of Cambridge HPC team, showed that they faced very similar issues, although their applications are very different.
The University of Cambridge used our Cobalt hardware design as the basis for their Wilkes general purpose cluster\footnote{\url{http://www.hpc.cam.ac.uk/services/wilkes.html}}, which reached \#2 on the November 2013 edition of the Green500\footnote{\url{https://www.top500.org/green500/}} list.
The size of this cluster made this decision particularly note-worthy.
It was a 128 node cluster, with just 8 nodes per rack, taking up 16 racks in total.
At 4U per node, this was not a particularly dense solution, but the, at the time, unique and abundant PCIe structure in these nodes was judged sufficiently desirable to justify the additional expense in terms of rack space.

Informal presentations of the Cobalt design to several industry partners, including senior Dell management, have resulted in an increased awareness of radio astronomy as an eScience.
It was difficult to find a chassis from any one of the major vendors that could meet the requirements of the Cobalt project.
It is hoped that our discussions with industry, using this project as an example, will improve the suitability of future HPC system designs for the next generation of radio telescopes.

The initial design approach taken in this project, where the hardware is closely matched to the software requirements, has since been successfully employed in the SKA Science Data Processor preliminary design~\citep{broekema:2015}.

\section{Acknowledgements}
The authors would like to thank the Center for Information Technology (CIT) at the University of Groningen and in particular Wietze Albers for his indispensable efforts in the design and prototyping phase and Hopko Meijering and Arjen Koers for their system and network administration during the development of Cobalt and its operational use.
We thank Dell Netherlands, and Patrick Wolffs in particular, for their close involvement in the prototyping phase of this project and their help in getting our custom cooling solution supported and certified.
We would like to express our appreciation to the anonymous reviewers for their detailed comments on an earlier version of this manuscript.

ASTRON is an institute of NWO, the Netherlands Organisation for Scientific Research.

\bibliographystyle{unsrt}
\bibliography{cobalt}
\end{document}